\documentstyle[pslatex]{article}
\RequirePackage{times}
\RequirePackage{courier}
\RequirePackage{mathptm}
\newtheorem{theorem}{Theorem}[section]
\newtheorem{proposition}[theorem]{Proposition}
\newtheorem{definition}[theorem]{Definition}
\newtheorem{example}[theorem]{Example}
\newtheorem{remark}[theorem]{Remark}

\begin{document}
\title{Partition Logics, Orthoalgebras and Automata}
\author{Anatolij DVURE\v{C}ENSKIJ$^1$, Sylvia PULMANNOV\'A$^1$ and Karl SVOZIL$^2$}
\date{}
\maketitle
\begin{center}  \footnote{ The paper has been partially
supported by the grant G 229/94 SAV, Bratislava, Slovakia,
and by the Mitteln zur F\"orderung der Auslandsbeziehungen an der
Technischen Universit\"at Wien.} Mathematical Institute,  Slovak Academy
of Sciences,\\ \v Stef\'anikova
49, SK-814 73 Bratislava, Slovakia\\
e-mail: anatolij.dvurecenskij@mat.savba.sk \qquad silvia.pulmannova@mat.savba.sk\\
$^2$Institute for Theoretical Physics, Technical University of Vienna\\
Wiedner Hauptstra\ss e 8--10/136, A-1040 Vienna, Austria\\
e-mail: svozil@tuwien.ac.at
\end{center}


\begin{abstract}
We investigate the orthoalgebras of certain
non-Boolean models which have a classical realization.
Our particular concern will be the
partition logics arising from the investigation of the empirical
propositional structure of Moore and Mealy type automata.
\end{abstract}

\renewcommand{\baselinestretch}{1.5}
\section{Introduction}

The investigation of classical models for non-Boolean algebraic
structures has brought up several interesting examples. Among them are
Cohen's ``firefly-in-a-box'' model \cite{Coh}, Wright's urn model
\cite{Wri},
as well as Aerts' vessel model \cite{Aerts}  featuring stronger-than
quantum
correlations. Another type of classical objects are automata
models, one of which has been introduced by Moore
\cite{Moo} in an attempt to model quantum complementarity in the context
of
effective computation.
 D. Finkelstein and S.R. Finkelstein
\cite{FiFi}, and subsequently Grib and Zapatrin
\cite{GrZa,GrZap}
 investigated the
propositional structure of certain automaton models by lattice
theoretical methods.
Svozil \cite{Svo} and Schaller and Svozil
\cite{ScSv1,ScSv2,ScSv3} introduced partition logics,
which appear to be a natural framework for the study of the
propositional structure of Moore and Mealy type automata.
Thereby, the set of automaton states is partitioned with respect to
identifiability in input/output experiments; and the single partitions
corresponding to Boolean algebras are pasted together to form more
general structures.

We describe here how non-classical propositional structures, in
particular partition logics of automata, fit into the scheme of
orthoalgebras.

\section{Boolean Atlases}

According to Lock and Hardegree \cite{LH1, LH2}, 
we consider a family of Boolean
algebras,  a Boolean atlas, which will be equivalent to quasi
orthoalgebras. Many
considerations
about co-measurable quantum propositional structures
deal with Boolean
subalgebras.
In addition, they are intuitively better understandable
than general quantum propositional logics.

A family $\cal {B}$ $ = \{B_i: i \in I\}$ of Boolean algebras is called a {\it Boolean
atlas} if it satisfies the following conditions (here the operations in
$B_i$ are denoted by an index $i$):
\begin{enumerate}

\item[{\rm (i)}] if $B_i \subseteq B_j,$ then $B_i = B_j;$\vspace{-2mm}

\item[{\rm (ii)}]  if $a,b \in B_i \cap B_j,$ then $a \le_i b$ iff
$a\le_j b;$\vspace{-2mm}

\item[{\rm (iii)}] $1_i = 1_j = 1$ and $0_i = 0_j=0$ for all $i,j \in I;$
\vspace{-2mm}
\item[{\rm (iv)}] if $a \in B_i \cap B_j,$ then $a^{\bot_i} = a^{\bot_j}$
for all $i,j \in I;$\vspace{-2mm}

\item[{\rm (v)}] if $a,b \in B_i \cap B_j$ and if $a \wedge_i b = 0_i,$ then
$a\vee_i b = a\vee_j b.$
\end{enumerate}

Note that $a,b \in B_i \cap B_j$ and yet $a \vee_i b
\ne a\vee_j b$ and $a\wedge_i b \ne a\wedge_j b.$ We define a {\it Boolean
manifold} to be a Boolean atlas which satisfies the condition

\begin{enumerate}
\item[] if $a,b \in B_i \cap B_j,$  then $a \vee_i b = a\vee_j b$ and
$a\wedge_i b = a\wedge_j b.$
\end{enumerate}

Let $\cal {B}$ $ = \{B_i: i \in I\}$ be a Boolean atlas, $a,b \in \bigcup_{i \in I}
b_i$ and $S \subseteq \bigcup_{i\in I} B_i.$ Then we say that

\begin{enumerate}
\item[{\rm (i)}] $a,b$ are {\it compatible} if there is $i \in I$
and $a,b \in B_i;$\vspace{-2mm}

\item[{\rm (ii)}] $a,b$ are {\it orthogonal} if there is $i \in I$ such that
$a,b \in B_i$ and $a \wedge_i b = 0_i.$ A subset $S$ is called {\it pairwise
orthogonal} if $a,b $ are orthogonal for any $a,b \in S;$\vspace{-2mm}

\item[{\rm (iii)}]  $S$ is {\it jointly compatible} if there is $i \in I$ with
$S \subseteq B_i;$ $S$ is {\it pairwise compatible} if $a,b $ are compatible
for any $a,b \in S;$\vspace{-2mm}

\item[{\rm (iv)}] $S$ is {\it jointly orthogonal} if there is $i \in I$
with $S \subseteq B_i$ and $S$ is pairwise orthogonal.
\end{enumerate}

\section{Orthoalgebras}

The notion of orthoalgebras  (or quasi orthoalgebras) goes back to
axiomatic models of quantum mechanics introduced by
Foulis and Randall
\cite{FR,RF}
as  special algebraic structures describing propositional logics.

A {\it quasi orthoalgebra} is a set $L$ endowed with two special
elements $0,1 \in L$ $(0 \ne 1)$ and equipped with a partially defined
binary operation $\oplus$ satisfying the following conditions for
all $a,b \in L:$

\begin{enumerate}
\item[{\rm (oai)}]  if $a\oplus b$ is defined, then $b \oplus a$ is
defined and $a \oplus b = b \oplus a$ (commutativity law);\vspace{-2mm}

\item[{\rm (oaii)}] $a \oplus 0$ is defined for any $a \in L$
and $a \oplus 0 = a;$\vspace{-2mm}

\item[{\rm (oaiii)}] for any $a \in L,$ there is a unique element
$a' \in L$ such that $a \oplus a'$ is defined and $a \oplus a' = 1$
(orthocomplementation law);\vspace{-2mm}

\item[{\rm (oaiv)}] if $a \oplus (a' \oplus b) $ is defined, then $ b = 0;$
\vspace{-2mm}

\item[{\rm (oav)}] if $a \oplus (a \oplus b) $ is defined, then
$a = 0;$\vspace{-2mm}

\item[{\rm (oavi)}] if $a \oplus b$ is defined, then $a \oplus (a \oplus b)'$
is defined and $b'= a \oplus (a \oplus b)'.$

\end{enumerate}

The following facts are true:

\begin{proposition} Let $L$ be a quasi orthoalgebra, $a, b \in L.$
Then
\begin{enumerate}
\item[{\rm (a)}] $0'=1,$ $1'=0;$\vspace{-2mm}
\item[{\rm (b)}] $(a')' = a;$\vspace{-2mm}
\item[{\rm (c)}] if $a \oplus b = a \oplus c,$ then $b = c;$\vspace{-2mm}
\item[{\rm (d)}] if $a \oplus b = 1,$ then $b = a'.$
\end{enumerate}
\end{proposition}

The unique element $a'$ is called {\it orthocomplement} of $a \in L,$
and the unary operation $':\, L \to L$ defined by $a \mapsto a',\, a \in L,$
is said to be an  {\it orthocomplementation}.
We shall say that two elements $a,b \in L$  (i) are {\it orthogonal},
and write $a \perp b,$ iff $a \oplus b $ is defined in $L$
(it is clear that $a \perp b$ iff $b \perp a$),
and (ii)
 $a \le b$ iff there is an element $c \in L$ with $a \oplus c = b.$

It is easily to  shown that the relation $\le$ is reflexive and
antisymmetric,
but needs not to be transitive. An associative quasi orthoalgebra,
i.e., a quasi orthoalgebra, for which the associative law
\begin{enumerate}
\item[{\rm (oavii)}] if $a \oplus b,\, (a\oplus b) \oplus c$ are defined
in $L,$ so are $b \oplus c$ and $a \oplus(b \oplus c),$ and $(a \oplus b)
\oplus c = a \oplus ( b \oplus c)$
\end{enumerate}
holds is said to be an {\it orthoalgebra}  (OA in abbreviation). In any
orthoalgebra, $\le$ is
transitive. On other hand it is possible to give an example of a quasi
orthoalgebra with transitive $\le$ which does not correspond to  any
orthoalgebra.

Due to Golfin \cite{Gol}, an orthoalgebra is a set $L$ with two special
elements $0,1 \in L$ $(0\ne 1)$ and endowed with a partial  binary
operation
$\oplus $
satisfying (oai), (oaiii), (oavii), and (oav*) if
$a \oplus a$ is defined, then $a = 0.$

The original idea of the partial binary operation $\oplus$  goes back
to Boole's pioneering paper \cite{Bool}, where he wrote $a + b$ as the logical
disjunction of events $a$ and $b$ when the logical conjunction $ab = 0,$
so that, for mutually excluding events $a$ and $b,$ $a + b$ is defined.
This is all that is needed for probability theory: if $ab = 0,$  then
$P(a+b) = P(a) + P(b).$ To avoid confusion, we write $a \oplus b$ for $a+b$
when $ab = 0.$

Note that one can rewrite axioms for a Boolean algebra in terms of
Boole's
ideas of $a+b$. For more details, see Foulis and Bennett \cite{FB}.

In addition, let $L$ be an orthomodular poset (OMP for abbreviation)
(or an orthomodular lattice, OML in short), i.e., a poset $L$ with the
least
and last elements $0$ and $1$ and a unary operation $^\bot :\, L \to L$,
called an {\it orthocomplementation}, such that, for all $a,b \in L$,
\begin{enumerate}
\item[{\rm (i)}]  $(a^\bot )^\bot = a;$\vspace{-2mm}
\item[{\rm (ii)}] if $a \le b,$ then $b^\bot \le a^\bot;$\vspace{-2mm}
\item[{\rm (iii)}] $a \vee a^\bot = 1;$\vspace{-2mm}
\item[{\rm (iv)}] if $a \le b^\bot$ (and we write $a \perp b$),
then $a \vee b \in L;$\vspace{-2mm}
\item[{\rm (v)}] if $a \le b,$ then $b = a \vee (a \vee b^\bot)^\bot$.
\end{enumerate}
(For OML, $L$ has to be additionally a lattice).
Then $L$ can be organized into an OA if the binary operation $\oplus$
is defined via $a \oplus b$ exists in $L$
iff $ a\le b^\bot$ and $a \oplus b :=
a \vee b$. The unary operation $':\, L \to L$ is defined via\
$a' := a^\bot, \, a \in L.$

We recall that if $L$ is an OA and $a,b \in L$ are mutually orthogonal, then
$a, b \le a \oplus b,$ and $ a\oplus b$ is the minimal upper bound for
$a$ and $b$ (i.e., $a,b \le a \oplus b,$ and if there is $ c \in L$ with
$a,b \le c \le a\oplus b,$ then $c = a\oplus b$), but this does not mean
that $a \vee b$ exists in $L,$ so that $L$ cannot be necessarily  an OMP.

A subset $A$ of a quasi OA (OA) $L$ is a {\it quasi suborthoalgebra}
({\it suborthoalgebra}) of  $L$ is (i) $0,1 \in A;$ (ii) if $a \in A,$
then $a' \in A;$ (iii) $a,b \in A$ with $ a \perp b$ implies
$a \oplus b \in L.$

If a (quasi) suborthoalgebra $A$ of $L$ is, in addition, a Boolean
algebra with respect to $\le $, $A$ is called a {\it Boolean
suborthoalgebra} of $L.$ Denote by $\vee_A$ and $\wedge_A$ the join and
the meet taken only in $A$, respectively. Then, $a \oplus b = a \vee_A
b$ whenever
$a,b
\in A$ and $L$ is an OA.  A maximal Boolean suborthoalgebra of $L$ is
called a {\it block.}

\section{Examples of Orthoalgebras}

We shall give a few examples of orthoalgebras
having classical physical interpretations.

\section*{ Firefly in a box}
According to Cohen \cite{Coh}, consider a
system consisting of a firefly in a box with a clear plastic window
at the front and another one on the side pictured in Figure
1.

\vspace{1cm}
\begin{center}
\unitlength 0.80mm
\linethickness{0.4pt}
\begin{picture}(56.00,47.00)
\put(0.00,17.00){\line(1,0){50.00}}
\put(50.00,17.00){\line(0,1){30.00}}
\put(50.00,47.00){\line(-1,0){50.00}}
\put(0.00,47.00){\line(0,-1){30.00}}
\put(25.00,17.00){\line(0,-1){2.00}}
\put(50.00,32.00){\line(1,0){2.00}}
\put(52.00,32.00){\line(0,1){0.00}}
\put(56.00,23.00){\makebox(0,0)[cc]{f}}
\put(56.00,38.00){\makebox(0,0)[cc]{b}}
\put(13.00,11.00){\makebox(0,0)[cc]{l}}
\put(37.00,11.00){\makebox(0,0)[cc]{r}}
\put(25.00,3.00){\makebox(0,0)[cc]{Fig. 1}}
\end{picture}
\end{center}
\vspace{1cm}

 Suppose each window has a thin vertical line drawn down the
center to divide the window in half. We shall consider two experiments
on the system: The experiment A: Look at the front window. The experiment
B: Look at the side window. The outcomes of A and B are: See a light
in the left half ($l_A,\, l_B$), right half ($r_A,\, r_B)$ of window
or see no light $(n_A, n_B)$. It is clear that $n_A = n_B =: n$
and we put $l_A = :l,\, r_A =: r,\, l_B =: f,\, r_B =: b$ $(f$
for the front, $b$ for the back).

The Greechie diagram of the corresponding propositional logic is
given by Figure 2. (Recall that here the small circles on one
smooth line denote mutually orthogonal atoms lying in the same block;
for more details on Greechie diagrams, see \cite{PtPu}.) The associated
Hasse diagram is given by Figure 3.

\vspace{1cm}

\begin{center}
\unitlength 1.00mm
\linethickness{0.4pt}
\begin{picture}(63.00,26.00)
\multiput(2.00,25.00)(0.36,-0.12){84}{\line(1,0){0.36}}
\multiput(32.00,15.00)(0.36,0.12){84}{\line(1,0){0.36}}
\put(62.00,25.00){\circle{0.00}}
\put(62.00,25.00){\circle{2.00}}
\put(47.00,20.00){\circle{2.00}}
\put(32.00,15.00){\circle{2.00}}
\put(17.00,20.00){\circle{2.00}}
\put(2.00,26.00){\circle{0.00}}
\put(2.00,25.00){\circle{2.00}}
\put(32.00,3.00){\makebox(0,0)[cc]{Fig. 2}}
\put(32.00,15.00){\circle*{2.00}}
\put(47.00,20.00){\circle*{2.00}}
\put(62.00,25.00){\circle*{2.00}}
\put(2.00,25.00){\circle*{2.00}}
\put(17.00,20.00){\circle*{2.00}}
\put(62.00,21.00){\makebox(0,0)[cc]{$b$}}
\put(47.00,16.00){\makebox(0,0)[cc]{$f$}}
\put(32.00,11.00){\makebox(0,0)[cc]{$n$}}
\put(17.00,16.00){\makebox(0,0)[cc]{$r$}}
\put(2.00,21.00){\makebox(0,0)[cc]{$l$}}
\end{picture}
\end{center}
\vspace{1cm}

\vspace{1cm}
\begin{center}
\unitlength 1.00mm
\linethickness{0.4pt}
\begin{picture}(63.00,81.00)
\put(2.00,35.00){\circle*{2.00}}
\put(17.00,35.00){\circle*{2.00}}
\put(32.00,35.00){\circle*{2.00}}
\put(47.00,35.00){\circle*{2.00}}
\put(62.00,35.00){\circle*{2.00}}
\put(62.00,55.00){\circle*{2.00}}
\put(47.00,55.00){\circle*{2.00}}
\put(32.00,55.00){\circle*{2.00}}
\put(17.00,55.00){\circle*{2.00}}
\put(2.00,55.00){\circle*{2.00}}
\put(32.00,75.00){\circle*{2.00}}
\put(32.00,15.00){\circle*{2.00}}
\multiput(32.00,35.00)(-0.12,0.16){126}{\line(0,1){0.16}}
\multiput(17.00,55.00)(-0.12,-0.16){126}{\line(0,-1){0.16}}
\multiput(2.00,35.00)(0.18,0.12){167}{\line(1,0){0.18}}
\multiput(32.00,55.00)(-0.12,-0.16){126}{\line(0,-1){0.16}}
\multiput(17.00,35.00)(-0.12,0.16){126}{\line(0,1){0.16}}
\multiput(2.00,55.00)(0.18,-0.12){167}{\line(1,0){0.18}}
\multiput(32.00,35.00)(0.12,0.16){126}{\line(0,1){0.16}}
\multiput(47.00,55.00)(0.12,-0.16){126}{\line(0,-1){0.16}}
\multiput(47.00,35.00)(0.12,0.16){126}{\line(0,1){0.16}}
\multiput(62.00,55.00)(-0.18,-0.12){167}{\line(-1,0){0.18}}
\put(32.00,35.00){\line(0,-1){20.00}}
\multiput(32.00,15.00)(0.12,0.16){126}{\line(0,1){0.16}}
\multiput(62.00,35.00)(-0.18,-0.12){167}{\line(-1,0){0.18}}
\multiput(32.00,15.00)(-0.12,0.16){126}{\line(0,1){0.16}}
\put(17.00,35.00){\line(0,1){0.00}}
\multiput(2.00,35.00)(0.18,-0.12){167}{\line(1,0){0.18}}
\multiput(2.00,55.00)(0.18,0.12){167}{\line(1,0){0.18}}
\multiput(32.00,75.00)(-0.12,-0.16){126}{\line(0,-1){0.16}}
\put(17.00,55.00){\line(0,1){0.00}}
\put(32.00,55.00){\line(0,1){20.00}}
\multiput(32.00,75.00)(0.12,-0.16){126}{\line(0,-1){0.16}}
\multiput(62.00,55.00)(-0.18,0.12){167}{\line(-1,0){0.18}}
\put(32.00,10.00){\makebox(0,0)[cc]{0}}
\put(32.00,2.00){\makebox(0,0)[cc]{Fig. 3}}
\put(2.00,30.00){\makebox(0,0)[cc]{$l$}}
\put(16.00,30.00){\makebox(0,0)[cc]{$r$}}
\put(29.00,30.00){\makebox(0,0)[cc]{$n$}}
\put(48.00,30.00){\makebox(0,0)[cc]{$f$}}
\put(62.00,30.00){\makebox(0,0)[cc]{$b$}}
\put(2.00,59.00){\makebox(0,0)[cc]{$l'$}}
\put(17.00,59.00){\makebox(0,0)[cc]{$r'$}}
\put(28.00,59.00){\makebox(0,0)[cc]{$n'$}}
\put(47.00,59.00){\makebox(0,0)[cc]{$f'$}}
\put(62.00,59.00){\makebox(0,0)[cc]{$b'$}}
\put(32.00,81.00){\makebox(0,0)[cc]{1}}
\put(32.00,55.00){\line(3,-4){15.00}}
\put(32.00,55.00){\line(3,-2){30.00}}
\end{picture}
\end{center}
\vspace{1cm}

A quantum mechanical realization of the above experiment has been given
by Foulis
and Randall \cite{FR}, Exam. III: Consider a device which, from time
to time, emits a particle and projects it along a linear scale.
We perform two experiments. Experiment A: We look to see if
there is a particle present. If there is not, we record the
outcome of A as the symbol $n.$ If there is, we measure its
position coordinate $x$. If $x\ge 1$, we record the outcome of $A$ as
the symbol $r$, otherwise we record the symbol $l.$ Similarly
for experiment B: If there is no particle, we record the outcome
of B as the symbol $n.$ If there is, we measure the $x$--component
$p_x$ of its momentum. If $p_x \ge 1,$ we write $b$  as for the outcome,
otherwise we write $f$. The propositional logic is the same as for
the firefly box system.

Another interesting model equivalent to the firefly box system
has been given by Wright \cite{Wri}.
It uses a generalized urn model.
Consider
an urn having balls which are all black except for one letter in
red paint and one letter in green paint, limited to one of the
five combinations of letters $r,l,n,f,b$ listed in Table 4.

\vspace{1cm}

\begin{center}
\begin{tabular}{|c|c|c|}\hline
Ball Type & Red & Green \\ \hline
1  & l & b \\
2  & l & f \\
3  & r & b \\
4  & r & f \\
5  & n & n \\
\hline
\end{tabular}

\begin{center} Tab. 4 \end{center}
\end{center}
There are the two experiments Red and Green. To execute the Red
experiment, draw a ball from the urn and examine it under a red
filter and record the letter you see. Note that under the red
filter, the green letter will appear black and will thus be invisible.
There are three outcomes $l,\,r,\,n.$ The Green experiment executes
using a green filter (all red letters will appear invisible).
The outcomes will be restricted to the letters $b,\, f,\, n,$ which
gives the propositional logic described by Figures 2 and 3.

\section*{Firefly in a three-chamber box}
Consider again a firefly,
but now
in a three-chamber box pictured in Figure 5.

\vspace{1cm}
\begin{center}
\unitlength 1.00mm
\linethickness{0.4pt}
\begin{picture}(41.17,45.00)
\put(0.17,10.00){\line(1,0){40.00}}
\put(20.17,45.00){\line(0,1){0.00}}
\put(20.17,45.00){\line(0,1){0.00}}
\multiput(20.17,45.00)(0.12,-0.21){167}{\line(0,-1){0.21}}
\put(40.17,10.00){\line(0,1){0.00}}
\multiput(20.17,45.00)(-0.12,-0.21){167}{\line(0,-1){0.21}}
\put(0.17,10.00){\line(0,1){0.00}}
\put(5.17,5.00){\makebox(0,0)[cc]{$l_A$}}
\put(34.17,5.00){\makebox(0,0)[cc]{$r_A$}}
\put(41.17,16.00){\makebox(0,0)[cc]{$l_C$}}
\put(27.17,40.00){\makebox(0,0)[cc]{$r_C$}}
\put(12.17,40.00){\makebox(0,0)[cc]{$l_B$}}
\put(-0.83,16.00){\makebox(0,0)[cc]{$r_B$}}
\put(20.17,-1.00){\makebox(0,0)[cc]{Fig. 5}}
\put(20.00,10.00){\line(0,1){5.00}}
\put(20.00,22.50){\line(0,-1){5.00}}
\put(26.00,25.50){\line(2,1){4.00}}
\put(13.83,25.50){\line(-2,1){3.67}}
\put(20.00,22.56){\line(-2,1){3.67}}
\put(20.00,22.56){\line(2,1){4.00}}
\end{picture}
\end{center}
\vspace{1cm}

The firefly is free to roam among the three chambers and to
light up to will. The sides of the box are windows with vertical
lines down their centers. We make three experiments, corresponding to
the three windows $A$, $B$ and $C$.
For each experiment $E$, we record $l_E,\, r_E, \, n_E$ if we see,
respectively,
a light to the left, right, of the center line or no light. It
is clear that we can identify $r_A =l_C =: e$, $r_C = l_B =: c$,
$r_B = l_A =: a$, but now we do not identify $f:= n_A,\, b := n_B,\,
d:= n_C.$

The propositional logic of this model has the Greechie diagram given
by Fig. 6 and the corresponding Hasse diagram by Fig. 7,

\vspace{1cm}
\begin{center}
\unitlength 1.00mm
\linethickness{0.4pt}
\begin{picture}(47.00,52.00)
\put(6.00,14.00){\line(1,0){40.00}}
\put(26.00,14.00){\circle*{2.00}}
\put(46.00,14.00){\circle*{2.00}}
\put(6.00,14.00){\circle*{2.00}}
\multiput(26.00,48.00)(0.12,-0.20){167}{\line(0,-1){0.20}}
\multiput(6.00,14.00)(0.12,0.20){167}{\line(0,1){0.20}}
\put(26.00,48.00){\circle*{2.00}}
\put(6.00,10.00){\makebox(0,0)[cc]{$a$}}
\put(11.00,33.00){\makebox(0,0)[cc]{$b$}}
\put(39.00,33.00){\makebox(0,0)[cc]{$d$}}
\put(26.00,52.00){\makebox(0,0)[cc]{$c$}}
\put(26.00,10.00){\makebox(0,0)[cc]{$f$}}
\put(46.00,10.00){\makebox(0,0)[cc]{$e$}}
\put(26.00,3.00){\makebox(0,0)[cc]{Fig. 6}}
\put(6.00,14.00){\line(1,0){40.00}}
\put(16.00,31.00){\circle*{2.00}}
\put(36.00,31.00){\circle*{2.00}}
\end{picture}
\end{center}

\vspace{1cm}
\begin{center}
\unitlength 1.00mm
\linethickness{0.4pt}
\begin{picture}(103.00,84.00)
\put(8.00,38.00){\circle*{2.00}}
\put(23.00,38.00){\circle*{2.00}}
\put(38.00,38.00){\circle*{2.00}}
\put(53.00,38.00){\circle*{2.00}}
\put(68.00,38.00){\circle*{2.00}}
\put(83.00,38.00){\circle*{2.00}}
\put(98.00,38.00){\circle*{2.00}}
\put(98.00,58.00){\circle*{0.00}}
\put(98.00,58.00){\circle*{2.00}}
\put(83.00,58.00){\circle*{2.00}}
\put(68.00,58.00){\circle*{2.00}}
\put(53.00,58.00){\circle*{2.00}}
\put(38.00,58.00){\circle*{2.00}}
\put(23.00,58.00){\circle*{2.00}}
\put(8.00,58.00){\circle*{2.00}}
\put(53.00,78.00){\circle*{2.00}}
\put(53.00,18.00){\circle*{2.00}}
\put(53.00,18.00){\line(0,1){20.00}}
\multiput(53.00,38.00)(-0.12,0.16){126}{\line(0,1){0.16}}
\multiput(38.00,58.00)(-0.12,-0.16){126}{\line(0,-1){0.16}}
\multiput(23.00,38.00)(-0.12,0.16){126}{\line(0,1){0.16}}
\multiput(8.00,58.00)(0.18,-0.12){167}{\line(1,0){0.18}}
\multiput(38.00,38.00)(-0.12,0.16){126}{\line(0,1){0.16}}
\multiput(23.00,58.00)(-0.12,-0.16){126}{\line(0,-1){0.16}}
\multiput(8.00,38.00)(0.18,0.12){167}{\line(1,0){0.18}}
\multiput(38.00,58.00)(0.12,0.16){126}{\line(0,1){0.16}}
\put(53.00,78.00){\line(0,-1){20.00}}
\multiput(53.00,58.00)(-0.12,-0.16){126}{\line(0,-1){0.16}}
\put(53.00,38.00){\line(0,1){0.00}}
\multiput(38.00,58.00)(0.18,-0.12){167}{\line(1,0){0.18}}
\multiput(68.00,38.00)(0.18,0.12){167}{\line(1,0){0.18}}
\multiput(98.00,58.00)(-0.12,-0.16){126}{\line(0,-1){0.16}}
\multiput(83.00,38.00)(-0.12,0.16){126}{\line(0,1){0.16}}
\multiput(68.00,58.00)(0.18,-0.12){167}{\line(1,0){0.18}}
\multiput(98.00,38.00)(-0.12,0.16){126}{\line(0,1){0.16}}
\multiput(83.00,58.00)(-0.12,-0.16){126}{\line(0,-1){0.16}}
\multiput(68.00,38.00)(-0.12,-0.16){126}{\line(0,-1){0.16}}
\multiput(53.00,18.00)(-0.12,0.16){126}{\line(0,1){0.16}}
\put(23.00,38.00){\vector(3,-2){30.00}}
\put(53.00,18.00){\vector(3,2){30.00}}
\multiput(98.00,38.00)(-0.27,-0.12){167}{\line(-1,0){0.27}}
\multiput(53.00,18.00)(-0.27,0.12){167}{\line(-1,0){0.27}}
\multiput(8.00,58.00)(0.27,0.12){167}{\line(1,0){0.27}}
\multiput(53.00,78.00)(-0.18,-0.12){167}{\line(-1,0){0.18}}
\multiput(68.00,58.00)(-0.12,0.16){126}{\line(0,1){0.16}}
\put(53.00,78.00){\line(0,1){0.00}}
\multiput(53.00,78.00)(0.18,-0.12){167}{\line(1,0){0.18}}
\put(83.00,58.00){\line(0,1){0.00}}
\multiput(98.00,58.00)(-0.27,0.12){167}{\line(-1,0){0.27}}
\put(53.00,78.00){\line(0,1){0.00}}
\multiput(53.00,58.00)(0.12,-0.16){126}{\line(0,-1){0.16}}
\put(3.00,34.00){\makebox(0,0)[cc]{$a$}}
\put(19.00,34.00){\makebox(0,0)[cc]{$b$}}
\put(34.00,34.00){\makebox(0,0)[cc]{$c$}}
\put(50.00,34.00){\makebox(0,0)[cc]{$d$}}
\put(62.00,34.00){\makebox(0,0)[cc]{$e$}}
\put(101.00,34.00){\makebox(0,0)[cc]{$a$}}
\put(53.00,12.00){\makebox(0,0)[cc]{0}}
\put(53.00,84.00){\makebox(0,0)[cc]{1}}
\put(2.00,54.00){\makebox(0,0)[cc]{$a'$}}
\put(18.00,55.00){\makebox(0,0)[cc]{$b'$}}
\put(32.00,56.00){\makebox(0,0)[cc]{$c'$}}
\put(48.00,56.00){\makebox(0,0)[cc]{$d'$}}
\put(63.00,56.00){\makebox(0,0)[cc]{$e'$}}
\put(103.00,55.00){\makebox(0,0)[cc]{$a'$}}
\put(53.00,3.00){\makebox(0,0)[cc]{Fig. 7}}
\put(87.00,56.00){\makebox(0,0)[cc]{$f'$}}
\put(87.00,35.00){\makebox(0,0)[cc]{$f$}}
\put(38.00,38.00){\line(3,2){30.00}}
\put(53.00,38.00){\line(3,4){15.00}}
\end{picture}
\end{center}
\vspace{1cm}
which is an orthoalgebra, called the Wright triangle,
 being no OMP. It is the most
 simple case of an OA which is not an OMP.
(Due to \cite{HNP},
an OA $L$ is not an OML iff it contains the Wright triangle as a suborthoalgebra
of $L$ in such a way that, for atoms $a,c,e$ of the corners of the triangle,
$a \oplus(c \oplus e)$ is not defined in $L.$)

In analogy with the generalized urn models, Wright \cite{Wri}, we can
describe the firefly three--chamber box system equivalently  as follows.
Consider an urn containing balls which are all black except for one
letter in red paint, one letter in green paint and one letter in blue
paint, limited to one of the following four combinations of letters
$a,\,b,\,c,\,d,\,e,\,f$ according to Table 8.
There are three experiments Red, Green
and Blue using  a red, green or blue filter. Assume now (somewhat
unphysically) that each one of
these three filters lets light through only in its own colour, and that
different colours  are invisible; i.e., they appear black.
The corresponding propositional logic is again
given by the Wright triangle.

\vspace{1cm}

\begin{center}
\begin{tabular}{|c|c|c|c|}\hline
Ball Type & Red & Green& Blue \\ \hline
1   & a & a & d\\
2   & c & f & c\\
3   & b & e & e\\
4   & b & f & d\\
\hline
\end{tabular}

\begin{center} Tab. 8  \end{center}
\end{center}

\section{Relations among Boolean Atlases and Quasi Orthoalgebras}

The following theorem has been proved by Lock and Hardegree and to be
self-contained we repeat their proof with small changes.

\begin{theorem}\label{th4.1} {\rm (1)} Every Boolean atlas defines a quasi orthoalgebra
in a natural way.

{\rm (2)} Every quasi orthoalgebra defines a Boolean atlas in a natural way.
\end{theorem}

{\bf Proof.} (1) Let ${\cal {B}} = \{B_i:\, i \in I\}$ be a Boolean
atlas. We define a quasi orthoalgebra $L$ as follows: $L := \bigcup_{i \in I} B_i,\,
0 = 0_i,\, 1 = 1_i,\, a' = a^{\bot_i}$ for any $i\in I$ such that
$a \in B_i.$ We say $a \perp b$ iff there is an $i \in I$ such that $a, b \in B_i$
and $a \wedge_i b = 0,$ and then $a \oplus b := a \vee_i b.$ The operations
are well-defined, and it can be shown that properties of quasi orthoalgebras
are satisfied.

We note that the relation $\le$ on $L$ is defined as follows:
$a \le b $ iff there is $x \in L$ with $ a \oplus x = b.$ This means the
following: there is an $i \in I$ with $a,x \in B_i,\, a \wedge_i x = 0,$
and $b = a \oplus x = a \vee_i x.$ This implies $a \le_i b.$ On the other hand,
if $a \le_i b$ for some $i \in I,$ then $a \wedge_i b^\bot = 0,$ where
$a \perp_i b^\bot.$ Therefore, $a \oplus b^\bot = a \vee_i b^\bot$ is defined, and,
moreover, $a \perp_i (a \oplus b^\bot)^\bot,$ so that $a \perp_i (a \oplus b^\bot)^\bot
= a \vee_i (a \vee_i b^\bot)^\bot = a \vee_i (a^\bot\wedge_i b) =b,$ whence $ a \le b.$

(2) Let $L$ be a quasi orthoalgebra. Let $ \{B_i:\, i \in I\}$ be
the set of all blocks of $L.$ Then ${\cal {B}} = \{B_i:\, i \in I\}$
is a Boolean atlas. \hfill $\Box$

\begin{example} Let $\Omega = \{1,2,3,4,5,6\}$ and let $B_1$ and $B_2$ be
 the Boolean algebras generated by $\{1\},$$ \{2\},$$ \{3\},$$
\{4\},$$
\{5,6\}$ and $\{1\},$$ \{2\},$$ \{3,4\},$$ \{5\},$$ \{6\},$ respectively
(with respect
the set-theoretic inclusion and $1_1 = 1_2= \Omega$). Then ${\cal {B}}
= \{B_1, B_2\}$ is a Boolean atlas, and $L = B_1 \cup B_2$ is,
according to Theorem {\rm \ref{th4.1}}, a quasi orthoalgebra. An easy
calculation shows that the order $\le $ induced by $\oplus$ in $L$
is not transitive. Indeed, we have $\{3\} \le \{3,4\},\, \{3,4\} \le
\{3,4,5\}$ but $\{3\} \not\le \{3,4,5\}$ although $\{3\} \subseteq \{3,4,5\},$
consequently, $L$ is not an OA.
\end{example}

\section{Partition Logics}

In this section, we present a notion of partition logics which
will have an intimate connection with special types of automata,
and which will generalize the results of Svozil
\cite{Svo}
and
 Schaller and Svozil
\cite{ScSv1, ScSv2, ScSv3}.

Let $L$ be a quasi orthoalgebra with  $\le.$ A non-void
subset $I$
of $L$ is said to be an {\it ideal} of $L$ if
\begin{enumerate}
\item[{\rm (i)}] if $a \in I, b \in L, b \le a,$  then $ b \in I;$    \vspace{-2mm}
\item[{\rm (ii)}]      $a,b \in I$ with $ a\perp b$ imply $a \oplus b \in I.$
\end{enumerate}
It is clear that $0 \in I.$
An ideal $I$ of $L$ is said to be (i)
{\it proper} if $I \ne L$ or, equivalently, $1 \not\in I;$ (ii) {\it prime}
if, for any $a \in L,$ either $a \in I$ or $a' \in I.$ We denote
by $P(L)$ the set of all prime ideals in $L.$

A {\it probability measure} (or also a {\it state}) on $L$ is a mapping $s:\, L \to [0,\,1]$ such that
(i) $s(1) = 1,$ and (ii) $s(a\oplus b) = s(a) + s(b)$ whenever $a \perp b.$
A probability measure $s$ is {\it two-valued} if $s(a) \in \{0,1\}$ for any $a \in L.$

We recall that there is a one-to-one correspondence between two-valued
probability measures and prime ideals: If $s$ is a two-valued probability measure, then $I_s = \{a \in L:
s(a) = 0\}$ is a prime ideal; and if $I$ is a prime ideal, then
$s_I: \, L\to [0,1]$ defined via $s_I(a) = 0 $ iff $a \in I,$ otherwise
$s_I(a) = 1,$ is a two-valued probability measure on $L.$

A set $\cal S$ of probability measures on $L$ is called {\it separating}
if for all $a,b \in L,\, a \ne b, $ there is a probability measure $s \in {\cal {S}}$
such that $s(a) \ne s(b).$  $L$ is called {\it prime} iff it has a separating
set of two-valued probability measure or, equivalently, for any different elements
$a,b \in L$ there is a prime ideal $I$ of $L$ such that $a \in I$ and
$b \not\in I.$

Let $\cal L$ be a family of  of quasi orthoalgebras (or OAs, OMP, Boolean
algebras, etc.) satisfying the following conditions: For all
$P,Q \in {\cal{L}},$ $P\cap Q$ is a quasi suborthoalgebra (subOA, sub OMP,
Boolean subalgebra, etc.) of both $P$ and $Q,$ and the partial orderings
and orthocomplementations coincide on $P \cap Q.$ Define the set
$L = \bigcup := \bigcup\{P:\, P \in {\cal{L}}\},$ a relation $\oplus $
and the unary operation $'$ as follows:
\begin{enumerate}
\item[{\rm (i)}]  $a \oplus b$ iff there is a $P \in {\cal {L}}$ such  that
$a,b \in P$ and $a \perp_P b,$ then $a \oplus b = a \oplus_P b;$  \vspace{-2mm}
\item[{\rm (ii)}]  $a'=b$ iff there is a $P \in  {\cal {P}}$ such that
$a,b \in P$ and $a^{'_P} = b.$
\end{enumerate}
The set $L$ with the above defined $\oplus$ is called the {\it pasting}
of the family $\cal L.$

Let $\cal R$ be a family of partitions of a fixed set  $M.$ The pasting
of the family of Boolean algebras $\{B_R:\, R \in {\cal {R}} \}$ is
called  {\it partition logic}, and we denote it as a couple
$(M, {\cal{R}}).$

\begin{remark} If $\cal {B}$ $ = \{B_i: i \in I\}$ is a Boolean atlas,
then $L = \bigcup_{i \in I} B_i$ with $\oplus$ and $'$ defined by
the last above {\rm (i)} and {\rm (ii)} is a pasting of a family of
Boolean algebras $\{B_i:\, i \in I\}.$
Moreover, $a \oplus b$ is defined iff $a,b \in B_i$ for some $i \in I$
with $a \wedge_i b = 0,$ and then $ a\oplus b = a \vee_i b.$
\end{remark}

We 
 recall that two quasi orthoalgebras $L_1$  and $L_2$ are {\it
isomorphic}
iff there is a one-to-one mapping $\phi:\, L_1 \to L_2$ such that
$a\oplus  b$ is defined in $L_1 $ iff $\phi(a) \oplus \phi(b)$ is
defined in $L_2$ and $\phi(a\oplus b) =\phi(a) \oplus \phi(b).$

\begin{theorem}\label{th5.2} A quasi orthoalgebra $L$ is
isomorphic to a partition logic if and only if $L$ is prime.
\end{theorem}

{\bf Proof.} (i)  Suppose that $L$ is isomorphic to a partition logic
$R = (M,{\cal {R}}).$ Without loss of generality, we may assume that $L=R.$
Take $A,B \in R$ such that $A \ne B$.  Then there is a point
$q \in (A\setminus B) \cup (B\setminus A).$ Put $P:= \{C \in R:\,
q \not\in C\}.$ Then $P$ is a prime ideal in $L.$ Indeed, let $C
\in P,$ and $D\le C.$ Then there is a partition $U \in {\cal
{R}}$ such that the Boolean algebra $B(U)$ generated by $U$
contains $D,C,$ and $D \le_{B(U)} C$ implies $D \subseteq C.$
It follows $q \not\in D,$ hence $D \in P.$

If $E,F \in R$ and $E
\perp F,$ there is a Boolean algebra $B(V)$ generated by a
partition $V$ such that  $E \cap_{B(V)} F = \emptyset.$ Moreover,
$E \oplus F = E \vee_{B(V)} F = E \cup F$ in $M.$ Therefore, $q \not\in
E \cup F,$ which gives $E \oplus F \in P.$

Finally, for every $C \in R,$ either $q \in C$ or $q \in M\setminus C,$
hence $P$ is a prime ideal.

(ii) Conversely, suppose that $L$ is prime. Let $M$ be the set
of all prime ideals in $L,$ i.e., $M = P(L).$ For $x \in L,$ we set
$p(x) :=\{P \in P(L):\, x \not\in P\}.$ Since $L$ is prime, the mapping
$p:\, L \to 2^M$ is injective. Moreover, $ x \perp y$ gives $p(x)
\cap p(y) =\emptyset$ and
$p(x\oplus y) = p(x) \cup p(y).$ Indeed, for any $P \in P(L),$
$x,y \in P$ iff
$x \oplus y \in P,$ consequently, $x \oplus y \not\in P $ iff either
$x \not\in P$ or $y \not\in P;$ since either $ x\in P$ or $y\in P$ for
any $P \in P(L)$ and all orthogonal elements $x$ and $y.$

In other words, we have proved that $x\perp y$ implies that the system
$R(x,y) := \{p(x), p(y),$ $ p((x\oplus y)')\}$ is a partition of $M.$
Let ${\cal{R}} = \{R(x,y):\, x,y \in L, x \perp y\}$ and let $R$
be the partition logic $(M,{\cal{R}}).$ For every $x \in L,$ $p(x)
\in R(x,x'),$ so that $p:\, L \to R$ is an injection, and by the
definition, also a surjection.

Let $A,B \in R$ with $A \perp_R B.$ That is, there is a partition
$P \in {\cal{R}}$ with $A,B \in B(P),$ and $A\wedge -{\cal {B(P)}} B =
\emptyset.$
By the definition of the partitions in $\cal R,$ there are elements
$x,y \in L$ such that $A = p(x), B= p(y)$ for some orthogonal
elements  $x,y \in L.$ This proves that $p$ is an isomorphism in question.
\hfill $\Box$

We say that two elements $a$ and $b$ of an OA $L$ have a {\it Mackey
decomposition} if there are three jointly orthogonal elements
$a_1, b_1, c$ in $L$ such that $a = a_1 \oplus c,\, b = b_1 \oplus
c.$ In OMPs any Mackey decomposition is unique, for OAs this is
not true, in general, however for prime orthoalgebras we have
the following result.

\begin{proposition}\label{prop5.3} A prime orthoalgebra has a unique
Mackey decomposition.
\end{proposition}

{\bf Proof.} Assume that $a$ and $b$ have two Mackey decompositions,
i.e., there are two jointly orthogonal systems $\{a_1, b_1, c_1\}
$ and $\{a_2, b_2, c_2\}$ such that $a = a_1 \oplus c_1 = a_2
\oplus c_2,$ $b= b_1 \oplus c_1 = b_2 \oplus c_2.$ Put $d_1 :=
(a_1 \oplus b_1 \oplus c_1)'$ and $d_2 := (a_2 \oplus b_2 \oplus
c_2)'.$ We assert that $d_1 = d_2.$

Assume the converse. Then there is a two--valued probability measure $s$ on
$L$ such that $s(d_1) = 1$ and $s(d_2) = 0.$ Hence, $s(a_1) = s(b_1)
=s(c_1) = 0,$ but one of $s(a_2), s(b_2), s(c_2) $ is 1. This
leads to
a contradiction, since $a_1 \oplus c_1 = a= a_2 \oplus c_2$ and
$b_1 \oplus c_1 = b = b_2 \oplus c_2.$ Therefore, $d_1 = d_2$,
and hence $a_1 \oplus b_1 \oplus c_1 = a_2 \oplus b_2 \oplus c_2.$
This entails $a \oplus b_1 = a \oplus b_2,$ so that $ b_1 = b_2$ and
$c_1 = c_2,$ consequently, $a_1 = a_2.$ \hfill$ \Box$

\section{Partition Logics and Automata logics}

Let an {\em alphabet} be a finite nonvoid set.
The elements of an alphabet are called {\em symbols}.
A {\em word} (or {\em string}) is a finite (possibly empty) sequence of
symbols.
The {\em length} of a word $w$, denoted by $|w|$, is the number of symbols
composing the string.
The {\em empty word} is denoted by $\epsilon$.
$\Sigma^*$ denotes the set of all words over an alphabet $\Sigma$.
The {\em concatenation} of two words is the word formed by writing
the first, followed by the second, with no intervening space.
Let $\Sigma$ be an alphabet.
$\Sigma^*$ with the concatenation as operation forms a monoid,
where the empty word $\epsilon$ is the identity.
A ({\em formal}) {\em language} over an alphabet $\Sigma$ is a
subset of $\Sigma^*$.

\begin{definition}
\label{moore}
A Moore automaton $M$ is a five-tuple
$M=(Q,\Sigma,\Delta,\delta,\lambda)$, where
\begin{enumerate}
\item[{\rm (i)}] $Q$ is a finite set, called the set of states;\vspace{-2mm}

\item[{\rm (ii)}]  $\Sigma$ is an alphabet, called the input
alphabet;\vspace{-2mm}

\item[{\rm (iii)}] $\Delta$ is an alphabet, called the output alphabet;
\vspace{-2mm}
\item[{\rm (iv)}]  $\delta$ is a mapping $Q \times \Sigma$ to $Q$,
called the transition function;
\vspace{-2mm}
\item[{\rm (v)}] $\lambda$ is a mapping $Q$ to $\Delta$, called the output function.
\end{enumerate}
\end{definition}
\begin{definition}
A Mealy automaton is a five-tuple $M=(Q,\Sigma,\Delta,\delta,\lambda),$
where $Q,\Sigma,\Delta,$ $\delta$ are as in the Moore automaton and
$\lambda$ is a mapping from $Q \times \Sigma$ to $\Delta$.
\end{definition}

Informally, a Moore automaton is in a state $q \in Q$,
emitting the output $\lambda(q) \in \Delta$ at any time.
If an input $a \in \Sigma$ is applied to the machine,
in the next discrete time step
the machine instantly assumes the state $p = \delta(q,a)$ and emits
the output $\lambda(p)$.
A Mealy machine emits the output at the instant of the transition
from one state to another, the output depending both on the previous
state and the input.

Suppose now an observer is performing experiments with a Moore or
Mealy automaton which is contained in a black box
with input-output interface.
Thus we are only allowed to observe
the input and output sequences associated with the box.
To conduct an experiment, the observer applies an input sequence
and notes the resulting output sequence.
Using this output sequence, the observer tries to interpret the
information contained in the sequence to determine the values of the
unknown parameters.

Suppose the observer conducts experiments on an automaton with a
known
transition table (i.e., the five-tuple $(Q,\Sigma,\Delta,\delta,\lambda)$)
but unknown initial state. This will be called the {\em initial state
identification problem.} Suppose further
that only a single copy of the
machine is available.

The logical structure of the initial-state identification problem can be
defined as follows. Let us call a proposition concerning the initial
state of the machine
{\em experimentally decidable} if there is an experiment $E$ which
determines the truth value of that proposition.
This can be done by performing $E$, i.e., by the input of a sequence of
input symbols $a_1,a_2,a_3,\ldots ,a_n$ associated with $E$, and by
observing the output sequence
\\
$\lambda_E(q)=\lambda(a_1,q), \ldots ,\lambda(\underbrace{\delta
(\cdots
\delta
(q,a_1)\cdots ,a_n)}_{n \mbox{ times}},a_n)$.
The most general form of a prediction concerning the
initial state $q$
of the machine is that the initial state $q$ is contained in a subset $P$ of the
state set $Q$.
Therefore, we may identify propositions concerning the initial state
with subsets of $Q$.
A subset $P$ of $Q$ is then  identified with the proposition that the
initial state is contained in $P$.

\begin{definition}
Let $E$ be an experiment (a preset or adaptive one), and let
$\lambda_E(q)$
denote the obtained output of an initial
state $q$.
$\lambda_E$ defines a mapping of $Q$ to the set of output sequences
$\Delta^*$. We define an equivalence relation on the state set $Q$ by \\
$q \stackrel{E}{\equiv} p$ iff $\lambda_E(q) = \lambda_E(p)$ \\
for any $q,p \in Q$.
We denote the partition of $Q$ corresponding to $\stackrel{E}{\equiv}$
by $Q/\stackrel{E}{\equiv}$.
Obviously, the propositions decidable by the experiment $E$ are
the elements of the Boolean algebra generated by $Q/\stackrel{E}{\equiv}$,
denoted by $B_E$.
There is also another way to construct the experimentally decidable
propositions of an experiment $E$.
Let $\lambda_E(P)  = \bigcup\limits_{q \in P}\lambda_E(q)$ be the direct
image of $P$ under $\lambda_E$ for any $P \subseteq Q$.
We denote the direct image of $Q$ by $O_E$, $O_E = \lambda_E(Q)$.

It follows that the most general form of a prediction concerning
the outcome $W$ of the experiment $E$ is that $W$ lays in a subset of
$O_E$.
Therefore, the experimentally decidable propositions consist of all
inverse images $\lambda_E^{-1}(S)$ of subsets $S$ of $O_E$,
a procedure which can be constructively formulated (e.g., as an
effectively computable algorithm), and which also
leads to the Boolean algebra $B_E$.
Let ${\cal B}$ be the set of all Boolean algebras $B_E$.
We call the partition logic $R= (Q,{\cal B})$ an {\em automaton
propositional calculus.}
\end{definition}

\begin{proposition}
To every partition logic  $R$ there exists an automaton
$M$ such that
$R = R(M)$.
\end{proposition}
{\bf Proof.}
Let $R = (Q,{\cal R})$ be a partition
logic.
Every $P \in {\cal R}$ can be rewritten as an indexed family $P =
(P_i)_{i
\in I_n},$ where the index set $I_n$ denotes the set $\{1,\ldots,n\}$ of
natural numbers.
We assume that $P_i \neq P_j$ for $i \neq j$.
$N$ denotes the greatest number of elements   in any partition $P \in
{\cal R}$. Let $M = (Q,{\cal R},I_N,\delta,\lambda)$ denote the
automaton corresponding to the partition logic $R = (Q,{\cal R})$.
What remains to be defined are the
transition  function   $\delta$ and the output function $\lambda$.
Let $p$ be an arbitrary element of $Q$.
Then, for all $q \in Q$ and for all $P \in  {\cal R}$, let
(i) $\delta(q,P) = p$ and
(ii) $\lambda(q,P) = i$ iff $q \in P_i$.

\section{Partition Logics in Examples}

\begin{example}\label{ex7.1} A ``Fano plane" pictured at Fig. 9 is
not a partition logic (it is not prime, it has only unique $s$
probability measure, namely, $s(x) = 1/3$ for any atom $x \in L.$
\end{example}

\vspace{1cm}
\begin{center}
\unitlength 1.00mm
\linethickness{0.4pt}
\begin{picture}(28.00,30.44)
\put(15.00,16.00){\circle*{2.00}}
\put(15.00,16.00){\circle{14.00}}
\put(15.00,9.00){\circle*{2.00}}
\put(15.00,9.00){\line(0,1){21.00}}
\multiput(15.00,30.00)(0.12,-0.21){100}{\line(0,-1){0.21}}
\multiput(15.00,30.00)(-0.12,-0.21){100}{\line(0,-1){0.21}}
\put(3.00,9.00){\line(1,0){24.00}}
\put(27.00,9.00){\circle*{2.00}}
\put(3.00,9.00){\circle*{2.00}}
\put(15.00,29.44){\circle*{2.00}}
\multiput(3.00,9.00)(0.21,0.12){84}{\line(1,0){0.21}}
\put(21.11,19.11){\line(0,1){0.00}}
\multiput(27.00,9.00)(-0.21,0.12){84}{\line(-1,0){0.21}}
\put(8.89,19.11){\line(0,1){0.00}}
\put(8.89,19.11){\line(0,1){0.00}}
\put(8.89,19.11){\circle*{2.00}}
\put(21.11,19.11){\circle*{2.00}}
\put(15.00,3.00){\makebox(0,0)[cc]{Fig. 9}}
\end{picture}
\end{center}
\vspace{1cm}

\begin{example}\label{ex7.2} The Wright triangle, pictured by
Fig. 6, is a partition logic. It has a separating set of
two-valued probability measures given by Table 10.

\vspace{1cm}
{\rm
\begin{center}
\begin{tabular}{|c|c|c|c|c|c|c|}\hline
measure & $a$ & $b$ & $c$ & $d$ & $e$ & $f$  \\ \hline
1     &  1  &  0  &  0  &  1  &  0  &  0 \\
2     &  0  &  0  &  1  &  0  &  0  &  1 \\
3     &  0  &  1  &  0  &  0  &  1  &  0 \\
4     &  0  &  1  &  0  &  1  &  0  &  1 \\
\hline
\end{tabular}

\begin{center} Tab. 10  \end{center}
\end{center}
}
 It is isomorphic to the following
partition logic given by $\Omega =\{1,2,3,4\}$ and three
decompositions of $\Omega$:
\\ 
 $\{\{1\},\, \{2\},\,\{3,4\}\},$
$\{\{2\},\, \{3\},\, \{
1
,4\}\}$ and $\{\{1\},\, \{3\}, \{2,4\}\}.$
The transition and output table of a Mealy automaton realizing the
Wright triangle is given by Table 11.

\vspace{1cm}
{\rm
\begin{center}
\begin{tabular}{|c|c|c|c|c|c|c|}\hline
$\delta$ & $1$ & $2$ & $3$ & $4$ \\ \hline
 $\{\{1\},\, \{2\},\,\{3,4\}\}$ &1&1&1&1\\
 $\{\{2\},\, \{3\},\,\{1,4\}\}$ &1&1&1&1\\
 $\{\{1\},\, \{3\},\,\{2,4\}\}$ &1&1&1&1\\
\hline
\hline
$\lambda$ & $1$ & $2$ & $3$ & $4$ \\ \hline
 $\{\{1\},\, \{2\},\,\{3,4\}\}$ &1&2&3&3\\
 $\{\{2\},\, \{3\},\,\{1,4\}\}$ &3&1&2&3\\
 $\{\{1\},\, \{3\},\,\{2,4\}\}$ &1&3&2&3\\
\hline
\end{tabular}

\begin{center} Tab. 11  \end{center}
\end{center}
}
We recall that according to {\rm \cite{Wri}}, it cannot be modeled in a
Hilbert space.
\end{example}

\begin{example}\label{ex7.3} An orthoalgebra given by Fig. 12 is
a partition logic.
Its system of all two valued probability measures is given in
Table 13.
A possible Mealy automaton realization is given in Table 14.

\vspace{1cm}
\begin{center}
{\rm
\unitlength 1.00mm
\linethickness{0.4pt}
\begin{picture}(79.00,64.00)
\put(43.00,19.00){\line(0,1){40.00}}
\multiput(43.00,59.00)(-0.21,-0.12){167}{\line(-1,0){0.21}}
\multiput(8.00,39.00)(0.21,-0.12){167}{\line(1,0){0.21}}
\multiput(43.00,19.00)(0.21,0.12){167}{\line(1,0){0.21}}
\multiput(78.00,39.00)(-0.21,0.12){167}{\line(-1,0){0.21}}
\put(43.00,59.00){\circle*{2.00}}
\put(43.00,39.00){\circle*{2.00}}
\put(43.00,19.00){\circle*{2.00}}
\put(61.00,29.00){\circle*{2.00}}
\put(78.00,39.00){\circle*{2.00}}
\put(61.00,49.00){\circle*{2.00}}
\put(26.00,49.00){\circle*{2.00}}
\put(8.00,39.00){\circle*{2.00}}
\put(26.00,29.00){\circle*{2.00}}
\put(2.00,39.00){\makebox(0,0)[cc]{$a$}}
\put(37.00,39.00){\makebox(0,0)[cc]{$d$}}
\put(61.00,24.00){\makebox(0,0)[cc]{$g$}}
\put(78.00,33.00){\makebox(0,0)[cc]{$h$}}
\put(61.00,55.00){\makebox(0,0)[cc]{$i$}}
\put(26.00,54.00){\makebox(0,0)[cc]{$b$}}
\put(43.00,64.00){\makebox(0,0)[cc]{$c$}}
\put(43.00,12.00){\makebox(0,0)[cc]{$e$}}
\put(26.00,24.00){\makebox(0,0)[cc]{$f$}}
\put(43.00,2.00){\makebox(0,0)[cc]{Fig. 12}}
\end{picture}
}
\end{center}
\vspace{1cm}

The corresponding decompositions of $\Omega =\{1,2,3,4,5,6\}$ are
$\{\{1,2\},\{
3,4,6\},$ $ \{5\}\}$
 for the block $a,b,c$,
\\ 
$\{\{5\},\{1,2,3,4\},\{6\}\}$
 for $c,d,e$,
$\{\{1,2\},\{3,4,5\},\{6\}\}$
for $a,e,f$,
 $\{\{6\},\{1,3,5\},\{2,4\}\}$
 for $e,g,h,$
\\ 
$\{\{2,4\},\{1,3,6\},\{5\}\}$
 for $h,i,c$.
\end{example}
\vspace{1cm}
{\rm
\begin{center}
\begin{tabular}{|c|c|c|c|c|c|c|c|c|c|}\hline
measure & $a$ & $b$ & $c$ & $d$ & $e$ & $f$ & $g$ & $h$ & $i$ \\ \hline
1     &  1  &  0  &  0  &  1  &  0  &  0  &  1  &  0  &  1  \\
2     &  1  &  0  &  0  &  1  &  0  &  0  &  0  &  1  &  0  \\
3     &  0  &  1  &  0  &  1  &  0  &  1  &  1  &  0  &  1  \\
4     &  0  &  1  &  0  &  1  &  0  &  1  &  0  &  1  &  0  \\
5     &  0  &  0  &  1  &  0  &  0  &  1  &  1  &  0  &  0  \\
6     &  0  &  1  &  0  &  0  &  1  &  0  &  0  &  0  &  1  \\
\hline
\end{tabular}

\begin{center} Tab. 13  \end{center}
\end{center}
}

\vspace{1cm}
{\rm
\begin{center}
\begin{tabular}{|c|c|c|c|c|c|c|}\hline
$\delta$ & $1$ & $2$ & $3$ & $4$ & $5$ & $6$ \\ \hline
$\{\{1,2\},\{3,4,6\},\{5\}\}$  &1&1&1&1&1&1\\
$\{\{5\},\{1,2,3,4\},\{6\}\}$  &1&1&1&1&1&1\\
$\{\{1,2\},\{3,4,5\},\{6\}\}$  &1&1&1&1&1&1\\
$\{\{6\},\{1,3,5\},\{2,4\}\}$  &1&1&1&1&1&1\\
$\{\{2,4\},\{1,3,6\},\{5\}\}$  &1&1&1&1&1&1\\
\hline
\hline
$\lambda$ & $1$ & $2$ & $3$ & $4$ & $5$ & $6$ \\ \hline
$\{\{1,2\},\{3,4,6\},\{5\}\}$  &1&1&2&2&3&2\\
$\{\{5\},\{1,2,3,4\},\{6\}\}$  &2&2&2&2&1&3\\
$\{\{1,2\},\{3,4,5\},\{6\}\}$  &1&1&2&2&2&3\\
$\{\{6\},\{1,3,5\},\{2,4\}\}$  &2&3&2&3&2&1\\
$\{\{2,4\},\{1,3,6\},\{5\}\}$  &2&1&2&1&3&2\\
\hline
\end{tabular}

\begin{center} Tab. 14  \end{center}
\end{center}
}

\begin{example}\label{ex7.4} Orthoalgebras given by Fig. 15  and
Fig. 16 are
 partition logics.
\end{example}

\vspace{1cm}
\begin{center}
\unitlength 1.00mm
\linethickness{0.4pt}
\begin{picture}(62.00,44.00)
\put(7.00,13.00){\line(0,1){30.00}}
\multiput(7.00,43.00)(0.16,-0.12){126}{\line(1,0){0.16}}
\multiput(27.00,28.00)(-0.16,-0.12){126}{\line(-1,0){0.16}}
\put(7.00,13.00){\line(0,1){0.00}}
\multiput(27.00,28.00)(0.24,0.12){126}{\line(1,0){0.24}}
\put(57.00,43.00){\line(0,-1){30.00}}
\multiput(57.00,13.00)(-0.24,0.12){126}{\line(-1,0){0.24}}
\put(27.00,28.00){\line(0,1){0.00}}
\put(27.00,28.00){\circle*{2.00}}
\put(57.00,28.00){\circle*{2.00}}
\put(57.00,43.00){\circle*{2.00}}
\put(7.00,43.00){\circle*{2.00}}
\put(7.00,28.00){\circle*{2.00}}
\put(7.00,13.00){\circle*{2.00}}
\put(57.00,13.00){\circle*{2.00}}
\put(42.00,20.00){\circle*{2.00}}
\put(42.00,35.00){\circle*{2.00}}
\put(17.00,35.00){\circle*{2.00}}
\put(17.00,20.00){\circle*{2.00}}
\put(2.00,13.00){\makebox(0,0)[cc]{$a$}}
\put(2.00,28.00){\makebox(0,0)[cc]{$b$}}
\put(2.00,43.00){\makebox(0,0)[cc]{$c$}}
\put(21.00,35.00){\makebox(0,0)[cc]{$d$}}
\put(27.00,32.00){\makebox(0,0)[cc]{$e$}}
\put(21.00,18.00){\makebox(0,0)[cc]{$f$}}
\put(39.00,16.00){\makebox(0,0)[cc]{$g$}}
\put(54.00,9.00){\makebox(0,0)[cc]{$h$}}
\put(62.00,43.00){\makebox(0,0)[cc]{$j$}}
\put(39.00,38.00){\makebox(0,0)[cc]{$k$}}
\put(62.00,26.00){\makebox(0,0)[cc]{$i$}}
\put(26.00,2.00){\makebox(0,0)[cc]{Fig. 15}}
\end{picture}
\end{center}
\vspace{1cm}

\vspace{1cm}
\begin{center}
\unitlength 1.00mm
\linethickness{0.4pt}
\begin{picture}(83.00,55.00)
\put(42.00,14.00){\line(0,1){40.00}}
\multiput(42.00,54.00)(-0.21,-0.12){167}{\line(-1,0){0.21}}
\multiput(7.00,34.00)(0.21,-0.12){167}{\line(1,0){0.21}}
\multiput(42.00,34.00)(0.21,0.12){167}{\line(1,0){0.21}}
\put(77.00,54.00){\line(0,-1){40.00}}
\multiput(77.00,14.00)(-0.21,0.12){167}{\line(-1,0){0.21}}
\put(42.00,34.00){\line(0,1){0.00}}
\put(42.00,34.00){\circle*{2.00}}
\put(7.00,34.00){\circle*{2.00}}
\put(25.00,44.00){\circle*{2.00}}
\put(42.00,54.00){\circle*{2.00}}
\put(77.00,54.00){\circle*{2.00}}
\put(59.00,45.00){\circle*{0.00}}
\put(59.00,44.00){\circle*{2.00}}
\put(59.00,24.00){\circle*{2.00}}
\put(25.00,24.00){\circle*{2.00}}
\put(42.00,15.00){\circle*{0.00}}
\put(42.00,15.00){\circle*{2.00}}
\put(77.00,15.00){\circle*{0.00}}
\put(77.00,14.00){\circle*{2.00}}
\put(77.00,34.00){\circle*{2.00}}
\put(2.00,34.00){\makebox(0,0)[cc]{$a$}}
\put(20.00,44.00){\makebox(0,0)[cc]{$b$}}
\put(37.00,54.00){\makebox(0,0)[cc]{$c$}}
\put(37.00,34.00){\makebox(0,0)[cc]{$d$}}
\put(37.00,11.00){\makebox(0,0)[cc]{$e$}}
\put(20.00,19.00){\makebox(0,0)[cc]{$f$}}
\put(54.00,44.00){\makebox(0,0)[cc]{$g$}}
\put(72.00,54.00){\makebox(0,0)[cc]{$h$}}
\put(83.00,34.00){\makebox(0,0)[cc]{$i$}}
\put(72.00,11.00){\makebox(0,0)[cc]{$j$}}
\put(55.00,19.00){\makebox(0,0)[cc]{$k$}}
\put(41.00,3.00){\makebox(0,0)[cc]{Fig. 16}}
\end{picture}
\end{center}
\vspace{1cm}

We note that combining the Wright triangles we can obtain plenty
of orthoalgebras which are partition logics.

\section{Partition Test Spaces}

Foulis and Randall \cite{FR, RF}  gave a new mathematical foundation of an
operational probability theory and statistics based upon a generalization
of the conventional notion of a sample space in the sense of Kolmogorov
\cite{Kol}.

Let us recall briefly main notions of their approach according to \cite{FB1}:

Let $X$ be a non-void set, elements of $X$ are called {\it outcomes}. We say
that a pair $(X,\cal T)$ is a {\it test space} iff $\cal T$ is a non-empty
family of subsets of $X$ such that (i) for any $x \in X,$ there is a $T \in
\cal T$ containing $x,$ and (ii) if $S,T \in \cal T$ and $S \subseteq T,$
then $S = T.$

Any element of $\cal T$ is said to be a {\it test}. We say that a subsets
$G$ of $X$ is an {\it event} iff there is a test $T \in \cal T$ such
that $G \subseteq T.$ Let us denote the set of all events in $X$ by
${\cal E} = {\cal E}(X,{\cal T}).$ We say that two events $F$ and $G$ are (i)
{\it orthogonal} to each other, in symbols $F \perp
G,$ iff $F \cap G = \emptyset$ and
there is a test $T \in {\cal T}$ such that $F \cup G \subseteq T;$
 (ii)  {\it local complements} of each other, in symbols $F\, loc\,G,$
iff $F \perp G$ and
there is a test $T \in {\cal T}$ such that $F  \cup G = T;$
 (iii) {\it perspective with axis} $H$
iff they share a common local complement $H.$ We write $F \approx_H G$ or
$F \approx G$ if the axis is not emphasized.

The test space $(X,\cal T)$ is {\it algebraic}
iff, for $F,G,H \in \cal E,$ $F\approx G$ and $F \, loc \, H$ entail $G \, loc\, H.$
Then $\approx $ is the relation of an equivalence, and, for any $A \in{\cal
E}(X,{\cal T}),$ we put $\pi(A) :=\{B \in {\cal E}(X,{\cal T}): \, B \approx A\}.$
Then $\Pi(X) :=\{\pi(A):\, A \in {\cal E}(X,{\cal T})\}$ is an orthoalgebra
\cite{FB1}.

Conversely, for any orthoalgebra $L,$ there is an algebraic test space $(X,{\cal T})$
such that $\Pi(X)$ is isomorphic with $L,$  \cite{FB1, Gud}.

For example, if $X$ is a unit sphere of a Hilbert space $H,$ then
$(X, {\cal B}(H)),$ where ${\cal B}(H)$ is the system of all orthonormal
bases in $H,$ is an algebraic test space, such that $\Pi(X)$ is
isomorphic to the complete OML $L(H)$ consisting of all closed subspaces
of $H.$

Let $(X,{\cal T})$ be a test space. A {\it weight} on $X$ is a function
$\omega:\, X \to [0,1]$ such that, for every $T \in {\cal T}$

\[ \omega(T) := \sum_{x \in T} \omega(x) = 1.\]
A weight $\omega$ is two-valued if $\omega(x) \in \{0,1\}$ for
any $x \in T$ and any $T \in {\cal T}.$ A set $\Delta$ of weights
on $X$ is {\it separating} if, for every $x_1, x_2 \in X,\, x_1
\ne x_2,$ there is a weight $\omega$ on $X$ such that
$\omega(x_1) \ne \omega(x_2).$

We concentrate now on the relationship between partition logics
with a special type of test spaces.

Let $X$ be a non-void set and $Y$ a non-void family of subsets
of a set $X.$ A couple $(Y,{\cal T}),$ where $ {\cal T}
\subseteq  2^Y,$ is said to be a {\it partition test space} of
$X$ if
\begin{enumerate}
\item[{\rm (i)}]    Every $T \in {\cal T}$ is a partition of $X$;  \vspace{-2mm}
\item[{\rm (ii)}]      For every $y \in Y,$ there is a $T \in
{\cal T}$ such that $y \in T.$
\end{enumerate}

\begin{proposition}\label{prop8.1} A partition test space is  a
test space.
\end{proposition}

{\bf Proof.} We have to show that if $T_1 \subseteq T_2,$ for
$T_1, T_2 \in {\cal T},$ then $T_1 = T_2.$ It follows from the
fact that $T_1$ and $T_2$ are partitions  of $X$.
\hfill $\Box$

\begin{proposition}\label{prop8.2} Let $(Y,{\cal T})$ be a partition
test space for $X.$ If $E, F \in {\cal E}(Y,{\cal T})$ and $E
\approx F,$ then $\bigcup E = \bigcup F.$
\end{proposition}

{\bf Proof.}  Let $G$ be a common complement of $E$ and $F.$ Then
$x \in \bigcup E$ iff $x \not\in \bigcup G$ iff $x \in \bigcup F.$
\hfill $\Box$

\begin{proposition} \label{prop8.3} A partition test space
$(Y,{\cal T})$ of $X$ is algebraic if every partition of $X$ consisting
of elements of $Y$ belongs to $\cal T.$
\end{proposition}

{\bf Proof.} Let $E,F,G,H$ be events such  that $E \approx_G F$
and $F\, loc\, H.$

For an event $E,$ put $\bigcup E:= \{x \in X:\ x \in y,\ y \in
E\}.$ From $E \approx_G F$ we obtain, for $x \in X,$ $x \in
\bigcup E$ iff $x \not\in \bigcup G$ iff $x \in \bigcup F,$ and from
$F\, loc\ H$ we obtain $x \in \bigcup F$ implies $x \not\in
\bigcup H.$

From this it follows that $F \cup H$ is a partition of $X$ and
so $F \cup H \in {\cal T}.$
\hfill $\Box$

Proposition \ref{prop8.2} implies that every partition test
space $(Y,{\cal T})$ of $X$ can be enlarged to an algebraic
partition test
space $(Y, {\cal U}),$ where ${\cal T} \subseteq {\cal U},$ and
$\cal U$
contains all partitions of $X$ which consist of elements of
$Y.$ The partition test space $(Y,{\cal U})$ with the latter
property will be called a {\it completion} of $(Y,{\cal T}).$ If
$\cal T$ and $\cal U$ coincide, we say that $(Y,{\cal T})$ is
{\it complete.}

If $(Y,{\cal T})$ is a complete partition test space, then for
any events $E,F$ with $\bigcup E = \bigcup F$ we have $E \approx F.$
Indeed, let $\bigcup E = \bigcup F,$ and let $G$ be any local
complement of $E.$ Then $\bigcup G = (\bigcup E)^c = X \setminus
\bigcup E = X \setminus \bigcup F,$ hence $G$ is also a local
complement of $F.$

\begin{proposition}\label{prop8.4} Let $(Y,{\cal T})$ be a
partition test space of the set $X.$ Then
\begin{enumerate}
\item[{\rm (i)}]  $\Pi(Y)$ is an OMP if $E,F,G \in {\cal E}(Y)$
with $E\perp F, \, F \perp G,\, G \perp E$ imply $(E \cup F)
\perp G.$
\item[{\rm (ii)}]   $\Pi(Y)$ is a concrete OMP\footnote{An OMP
$L$ is a {\it concrete logic} if it is isomorphic to a family
$ \cal L$ of subsets of a set $\Omega$ such that (i) $\Omega \in
{\cal L}.;$ (ii) If $A,B \in {\cal L}$ and $A \cap B =
\emptyset,$ then $A \cup B \in {\cal L}.$}
 if $(\bigcup E_1) \cap (\bigcup E_2) = \emptyset $ iff $E_1
\perp E_2.$
\end{enumerate}
\end{proposition}

{\bf Proof.} (i) It is evident.

(ii) According to Proposition \ref{prop8.2}, $\pi(E)$ can be identified
with $\bigcup E \subseteq X.$
\hfill $\Box$

\begin{remark}\label{rem8.5} The same set $L$ can be the logic
of several partition test spaces. A concrete logic $L$ can have
a test space not satisfying the condition (ii). Indeed, let $X =
\{1,2,3,4\}$ and take $(Y,{\cal T}),$ where $Y = \{\{1\},
\{3,4\}, \{2\}, \{2,4\}, \{3\}\},$ ${\cal T} = \{T_1, T_2\}$ and
$T_1 = \{\{1\}, \{3,4\}, \{2\}\},$ $T_2 =
\{\{1\}, \{2,4\}, \{3\}\}.$ Then $\Pi(Y)$ is a concrete OMP
(it is isomorphic to Fig. 2) with $\{2\} \cap \{3\} = \emptyset$,
but $\{2\} \not\perp \{3\}.$
\end{remark}

\begin{theorem}\label{thm8.6} A test space $(X,{\cal T})$ is isomorphic
to a partition test space if and only if it possesses a
separating family of two--valued weights.
\end{theorem}

{\bf Proof.} Let $(Y,{\cal T})$ be a partition test space of
$X.$ If $y_1, y_2 \in Y,\, y_1 \ne y_2,$ then $(y_1 \setminus y_2)
\cup (y_2 \setminus y_1)$ possesses at least one point, say $x$,
of $X.$ Define a function $\omega:\, Y \to \{0,1\}$ by putting $\omega
(y) =1$ iff $x \in y,$  otherwise we put $\omega(y) = 0.$
Then $\omega$ is a two--valued weight on $(Y,{\cal T})$, and
$\omega(y_1) \ne \omega(y_2).$

Conversely, let $(X,{\cal T})$ be a test space with a separating
family $\Delta$ of two--valued weights. Define $\phi(x) :=
:=
\{\omega \in \Delta:\, \omega(x)= 1\}, x \in X,$ and $\phi(T) :=
\{\phi(x):\, x \in T\},\ T \in {\cal T}.$

Consider $(\phi (X),\, \phi({\cal T}))$, where $\phi(X) :=
\{\phi(x):\,
x \in X\}$ and $\phi({\cal T}) := \{\phi(T):\, T \in {\cal T}\}.$
We claim that $(\phi(X),\, \phi({\cal T}))$ is a partition
test space of $X$, where $\phi(X) \subseteq 2^{\Delta},\ \phi(T)$ is a
partition of $\Delta$ for any $T \in {\cal T}.$ Observe that,
for any $\omega \in \Delta,$ $\omega(T) = 1 = \sum_{x \in T} \omega(x),$
so that there is a point $x_0 \in T$ such that $\omega(x_0)=1$
and $\omega(x) = 0$ for any $x \ne x_0.$ That is, for any $\omega
\in \Delta$ and for any $T \in {\cal T},$ there is a unique $x
\in T$ such that $\omega \in \phi(x).$ This implies that every
$\phi(T)$ is a partition of $\Delta.$
\hfill $\Box$

\begin{theorem}\label{thm8.7} There is a one-to-one
correspondence (up to isomorphism) between partition logics and partition
test spaces.
\end{theorem}

{\bf Proof.} Let $(Y,{\cal T})$ be a partition test space for a
set $X.$ For any event $E \subseteq T,\ T \in {\cal T},$ define
$u(E) := \bigcup E.$ We have if $E \approx F,$ then $\bigcup E =
\bigcup F.$ Define $L:= \{ \bigcup E:\ E \in {\cal E}(Y)\}.$ For
every $T \in {\cal T},$ $u(T) := \{ u(E):\, E \subseteq T\}$ is a
Boolean algebra. Indeed, every $u(E)$ is a union of some sets
from the partition $T$ of $X.$ For $a,b \in L,$ define $a \perp
b$ iff there are disjoint $E,F \in {\cal E}(Y)$ with $E \cup F \subseteq T$
for some $T \in {\cal T},$ and $a = u(E),\, b = u(F)$; and define
$a\oplus b = u(E \cup F),\, a' =u(T \setminus E)$ when $a = u(E),\,
E \subseteq T \in {\cal T}.$ Clearly, $u(T) = X$ for every $T
\in {\cal T}$ is the greatest elements in $L$ (by the ordering
$a \le b$ iff $a \perp b'$). Clearly, $L$ is a pasting of Boolean
algebras $\{u(T):\, T \in {\cal T}\}.$ This $L$  will be called
the {\it logic} of $(Y, {\cal T})$ in $X.$

Conversely, if $L$ is a partition logic, that is, $L$ is a pasting
of Boolean algebras $B(T_i), \,i\in I,$ where $T_i$ is a
partition of a set $X \ne \emptyset$ for any $i \in I,$ then put
\[ Y = \bigcup_{i \in I} \{y :\, y \in T_i\}.\]
The couple $(Y,\{T_i:\, i \in I\})$ is a partition test space of
$X,$ and its logic is isomorphic with $L, $ and the proof is complete.
\hfill $\Box$

We recall that all examples in the previous section are arising
by the way described in Theorem \ref{thm8.6} and Theorem \ref{thm8.7}.

\section{Concluding remarks}

We have thus far established a relationship between quasi orthoalgebras,
partition test spaces and (automaton) partition logics. Thereby we have
made use of concepts and techniques used in the foundations of quantum
mechanics. These considerations may also have some relevance for the
intrinsic perception of computer-generated universes (in ``pop-science''
jargon: virtual realities), since the input-output analysis underlying
the automaton propositional calculus and thus partition logics are
exactly those structures which are recovered by investigating  those
universes with methods which are operational therein.

\end{document}